# Persistent Interfacial Topological Hall Effect Demonstrating Electrical Readout of Topological Spin Structures in Insulators

Jing Li,[1†] Huilin Lai,[2,3†] Andrew H. Comstock,[4†] Aeron McConnell,[4] Bharat Giri,[1] Yu Yun,[1] Tianhao Zhao,[5] Xiao Wang,[6] Yongseong Choi[7], Xuemei Cheng,[6] Jian Shen,[2,3] Zhigang Jiang,[5] Dali Sun,[4*] Wenbin Wang,[2,3*] and Xiaoshan Xu[1*]

[1]Department of Physics and Astronomy and Nebraska Center for Materials and Nanoscience, University of Nebraska, Lincoln, Nebraska 68588, USA

[2] State Key Laboratory of Surface Physics and Institute for Nanoelectronic Devices and Quantum Computing, Fudan University, Shanghai 200433, China

[3] Department of Physics, Fudan University, Shanghai 200433, China

[4]Department of Physics and Organic and Carbon Electronics Lab (ORaCEL),

North Carolina State University, Raleigh, North Carolina 27695, USA

[5]School of Physics, Georgia Institute of Technology, Atlanta, Georgia, 30332, USA

[6]Department of Physics, Bryn Mawr College, Bryn Mawr, Pennsylvania 19010, USA

[7]Advanced Photon Source, Argonne National Laboratory, Lemont, Illinois 60439, USA

†Joint first authors.

*Corresponding authors: dsun4@ncsu.edu, wangwb@fudan.edu.cn, xiaoshan.xu@unl.edu

## Abstract

Conventional topological Hall effects (THE) require conducting magnets, leaving insulating systems largely inaccessible. Here we introduce the interfacial topological Hall effect (ITHE), where the noncoplanar spin textures of insulating magnets are imprinted onto an adjacent heavy metal via the magnetic proximity effect (MPE) and detected electrically. In Pt/h-$LuFeO_3$ bilayers, h-$LuFeO_3$ hosts a topological spin structure robust against high magnetic fields, arising from a 120° triangular spin lattice with small spin canting that yields nontrivial topology but minimal magnetization. This generates a giant Hall response in Pt up to 0.5% of the longitudinal resistivity and a Hall-conductivity/magnetization ratio above 2 $V^{-1}$, clearly distinguishable from the spin Hall Hanle effect background. Field- and temperature-dependent analysis further reveals that Pt nanoclusters inherit topological textures from h-$LuFeO_3$ via MPE. Unlike the conventional THE narrow peak-and-dip features, ITHE in Pt/h-$LuFeO_3$ persists across a broad magnetic field range up to 14 T, demonstrating the exceptional stability of the underlying topological spin structure. This establishes ITHE as a powerful and sensitive probe for topological magnetism in ultrathin insulating films and paves the way for new spintronic applications.

## Introduction

Topological Hall effects (THE) arise when itinerant electrons traverse non-coplanar spin structures, acquiring a Berry phase that acts as an emergent magnetic field and deflects their motion.[1–3] In conventional systems, THE is typically limited to conducting magnets, leaving insulating materials largely inaccessible to electrical detection.

Interfacial topological Hall effect (ITHE) offers a promising route to extend topological transport to metal/insulator interfaces. Here, a heavy metal near the Stoner instability (e.g., Pt or Pd[4,5]) acquires a topological spin structure via magnetic proximity effect (MPE)[6,7] from an adjacent insulating layer with non-trivial spin topology (**Fig. 1a**). Because the emergent field depends on spin topology rather than net magnetization[8–11], ITHE can produce a sizable Hall signal even when the induced magnetization in the heavy metal is weak, enabling the electrical readout of topological spin textures in insulating magnets[12]. Clear demonstration of ITHE, however, remains elusive: MPE often induces only weak magnetization[4,13–17], and multiple interfacial spin-transport processes can obscure the signal[18–22]. The central scientific challenge is therefore to unambiguously identify a Hall response that originates from topological spin textures, establishing ITHE as a distinct and robust probe of magnetism in insulating systems.

To address this challenge, we studied Pt/h-$LuFeO_3$ bilayers. h-$LuFeO_3$ is a magnetic insulator that hosts a topological spin structure robust against high magnetic fields[23–27]. Its in-plane 120° triangular spin lattice with small spin canting along the c-axis produces a tiny net magnetization (~0.025 $\mu_B$/Fe, 3.9 kA/m)[23–26], yet a non-coplanar spin structure with non-trivial topology (**Fig. 1b**). The emergent field, proportional to the solid angle subtended by the spin trimer and inversely proportional to its base area, is estimated to reach approximately 50 tesla for h-$LuFeO_3$ (See Supplementary Materials **S1**). Crucially, unlike conventional topological spin structures, which often exist within a narrow magnetic field range[28], the robust non-coplanar spin configuration of h-$LuFeO_3$ provides a unique and stable platform to realize and probe ITHE across a broad magnetic-field range.

We show that Pt/h-$LuFeO_3$ exhibits a giant positive Hall signal reaching up to 0.5% of the longitudinal resistivity and a Hall-conductivity/magnetization ratio above 2 $V^{-1}$. This positive Hall signal vanishes near the Néel temperature ($T_N$) of h-$LuFeO_3$, is absent in Pt/h-$LuMnO_3$ (which lacks spin canting and hence topological spin structure[29]), and decays with Pt thickness, confirming its topological and interfacial origin. These results establish ITHE as a sensitive probe of topological magnetic order even in ultrathin insulating films. They also reveal that robust topological spin structures can generate a Hall signal across a broad field range, rather than showing the conventional peak-and-dip feature[28], providing a robust and versatile route for electrical detection of topological magnetism.

## ITHE with a giant Hall-conductivity/magnetization ratio

**Fig. 2a** shows the Hall resistivity $\rho_{xy}$ normalized by longitudinal resistivity $\rho_{xx}$ of a Pt (3.1 nm) /h-$LuFeO_3$ (15 nm) sample measured at different temperatures (see also **Fig. S1&S2**). At high fields, the total Hall signal can be expressed as $\rho_{xy} = \rho_{xy,OHE} + \rho_{xy,AHE} + \rho_{xy,SHHE} + \rho_{xy,ITHE}$, where $\rho_{xx,OHE}$, $\rho_{xx,AHE}$, $\rho_{xx,SHHE}$, and $\rho_{xx,ITHE}$ are the contributions from ordinary Hall effect, anomalous Hall effect (AHE), spin Hall Hanle effect (SHHE)[30], and ITHE, respectively. Our previous work indicates that at room temperature the Hall response of Pt/h-$LuFeO_3$ is dominated by OHE and SHHE, both negative in positive fields[31]. As the temperature decreases, however, a strong positive component emerges, eventually reversing the overall sign of $\rho_{xy}$ and becoming dominant at low temperature. This is consistent with the appearance of magnetic order below the $T_N$ (≈130 K) of h-$LuFeO_3$[32], indicating a magnetic origin.

The magnitude of this positive Hall response is striking, considering the small magnetization of h-$LuFeO_3$. At 25 K, $\rho_{xy}/\rho_{xx}(0)\approx4\times10^{-3}$ is over an order of magnitude larger than the corresponding values in Pt/$Y_3Fe_5O_{12}$ ($\rho_{xy}/\rho_{xx}(0)$~$1\times10^{-4}$)[18,22], despite h-$LuFeO_3$ having a magnetization nearly two orders of magnitude smaller (3.9 kA/m vs. 140 kA/m)[33]. Our x-ray magnetic circular dichroism (XMCD) measurements (**Fig. S3**) also indicate minimal induced Pt magnetization in Pt/h-$LuFeO_3$ compared with that in Pt/$Y_3Fe_5O_{12}$.[16] Using $\rho_{xy}/\rho_{xx}(0)\approx4\times10^{-3}$ and $\rho_{xx}\approx4.5\times10^{-7}$ Ω m (see **Fig. S1**), we obtain $\sigma_{xy}\approx1\times10^{-3}$ S/m. Even using the conservative upper bound $M$ = 3.9 kA/m (from h-$LuFeO_3$) for the induced Pt magnetization, we estimate a lower bound of $\sigma_{xy}/M\approx2$ $V^{-1}$, at least one order of magnitude larger than the typical anomalous Hall factor (~0.1 $V^{-1}$) [34], which makes a conventional AHE origin highly implausible.

Instead, all evidence points to ITHE as the origin of the positive Hall signal. First, no such contribution appears in Pt/$LuMnO_3$ (**Fig. 2b**, see also **Figs. S4&S5**), which shares the same crystal symmetry as h-$LuFeO_3$ but lacks spin canting[29], as illustrated in **Fig. 2c&d**. Second, Pt/h-$YbFeO_3$ bilayers, in which Fe spins cant similarly but the rare-earth sublattice differs[35,36] in comparison to h-$LuFeO_3$, show a similar magneto transport behavior (see **Figs. S6&S7**). Third, using the emerging field from the topological spin structure of h-$LuFeO_3$ ($B_e\approx50$ tesla, See Supplementary Materials **S1**) and the mobility of Pt (≈ 1 $cm^2$/Vs)[30], we can estimate $\rho_{xy}/\rho_{xx}\approx5\times10^{-3}$ which is close to the observation in **Fig. 2a**. Notably, ITHE here does not display the peak-and-dip feature of conventional THE, which arises from the instability of topological spin structures under high magnetic fields[28]. In contrast, the robust topological spin structure due to the strong intra-layer interaction in h-$LuFeO_3$ [25], allows ITHE to persist in a broad range of magnetic fields.

## Separating ISHE from background SHHE

Because the Hall signal in **Fig. 2a** contains contributions from both SHHE and ITHE, extracting ISHE contributions requires a quantitative measurement of $\rho_{xy,SHHE}$. This can be achieved using the established relationship between the SHHE Hall effect $\rho_{xy,SHHE}$ and its magnetoresistance counterpart $MR_{SHHE}$, where $MR\equiv\Delta\rho_{xx}(B)/\rho_{xx}(0)$ and $\Delta\rho_{xx}(B)\equiv\rho_{xx}(B)-\rho_{xx}(0)$, $B$ is the magnetic field[30].

As illustrated in **Fig. S2a**, SHHE arises when a spin current generated by the spin Hall effect is reflected at the metal/insulator interface, with precession in an applied magnetic field leading to both negative $\rho_{xy,SHHE}$ and positive $MR_{SHHE}$. When the field is applied along the $y$ direction, $MR_{SHHE}$ vanishes due to the absence of precession[30], so the $MR$ anisotropy $MR(B_z)$-$MR(B_y)$=[($\rho_{xx}(B_z)$-$\rho_{xx}(B_y)$)]/$\rho_{xx}(0)$ can be used to characterize SHHE. Indeed, the data can be well fit using SHHE, as shown in **Fig. S2d**. The extracted temperature dependence of $\theta_{SH}$ and $\tau_s$ (**Fig. S2e**) closely resembles that of Pt/h-$LuMnO_3$ in **Fig. S5c** and that of Pt/h-$YbFeO_3$ in **Fig. S7b**, except that $\theta_{SH}$ is larger in Pt/h-$LuFeO_3$, as observed previously[31].

With SHHE parameters determined (**Fig. S2e**), one can calculate the SHHE contribution to the Hall effect $\rho_{xy,SHHE}$, and isolate the ITHE contribution as $\rho_{xy,ITHE}=\rho_{xy}-\rho_{xy,SHHE}$ (**Fig. 3a**). The successful separation of negative SHHE and positive ITHE suggests that the sign reversal in **Fig. 2a**, also widely reported in other Pt/magnetic insulator systems[18–22,37], emerges naturally from the competition between these two effects.

## ITHE from superparamagnetic clusters

The field and temperature dependence of $\rho_{xy,ITHE}$ in **Fig. 3a** resembles a "superparamagnetism"-like response[21] from magnetic nanoclusters (illustrated in **Fig. 3b**):

$$M \propto \mu \left(\coth x - \frac{1}{x}\right) \qquad (1)$$

where $x \equiv \frac{\mu B}{k_B T}$, $\mu$ is the magnetic moment of the cluster, $T$ is temperature, $k_B$ is the Boltzmann constant. For topological spin structures in h-$LuFeO_3$, magnetization $M$ is proportional to spin canting, which in turn is proportional to the emergent field $B_e$. Consequently, ITHE $\rho_{xy,ITHE}$, which is proportional to $B_e$, should track $M$. Indeed, the Hall effect in **Fig. 3a** can be well fit using Eq. (1) across all measured temperatures. Combining SHHE and ITHE, the full Hall response $\rho_{xy}/\rho_{xx}(0)$ is quantitatively reproduced, as shown in **Fig. 2a**.

The superparamagnetic-like behavior suggests that a thin interfacial layer of Pt acquires magnetic order via MPE in the form of nanoclusters carrying topological spin structures. The persistent ITHE at high field confirms that the topological spin structure within each nanocluster is robust, consistent with the strong intra-layer interaction[25]. These observations also emphasize that the conventional peak-and-dip feature of THE arises from the field-induced instability of topological spin structures[28], not necessarily intrinsic to topological spin structures. In fact, ITHE in Pt/h-$LuFeO_3$ persists up to 14 T (**Fig. S8**), the maximum field accessible in our measurements. This exceptional stability also highlights h-$LuFeO_3$ as a unique and stable platform to investigate field-resilient topological spin structure and THE.

For $T$<100 K, nonlinear ITHE in **Fig. 3a** allows determination of $\mu = 34 \pm 12\ \mu_B$. In the low field (linear) limit, Eq. (1) reduces to $M \propto \frac{\mu^2 B}{T}$, implying $\frac{M}{B} T \propto \mu^2$. We then use the low-field slope in **Fig. 3a** times temperature $\frac{d(\rho_{xy}/\rho_{xx}(0))}{dB} T$, which is expected to be proportional to $\mu^2$, as the parameter to characterize the magnetic order behind the ITHE. **Fig. 3c** shows that $\frac{d(\rho_{xy}/\rho_{xx}(0))}{dB} T$ decreases with temperature and vanishes at high temperature when the magnetic order in h-$LuFeO_3$ is lost.

The negative $MR(B_y)$ (see **Fig. S2b**) that correlates with the appearance of ITHE, can be understood by analogy with giant magnetoresistance (GMR) in magnetic alloys. In the two-current model[38] (See Supplementary Materials **S7**), spin-dependent scattering between magnetized Pt nanoclusters leads to $MR \propto -M^2$ or $(-MR)^{1/2} \propto |M|$. Plotting $[-MR(B_y)]^{1/2}$ (**Fig. 3d**) yields a curve closely resembling **Fig. 3a**, which can also be fit using Eq. (1). For $T$<100 K, we extract $\mu = 52 \pm 11\ \mu_B$, slightly larger than the value from Hall response. Importantly, the low-field slopes multiplied by temperature, taken as the order parameter, collapse onto the same trend in **Fig. 3c** for both Hall and MR data. This unified description reinforces the picture of Pt interfacial nanoclusters acquiring topological spin structures via MPE.

## ITHE for measuring topological spin orders in ultrathin insulators

**Fig. 3c** confirms that low-field slope*$T$ can be a unified parameter to describe the magnetic order behind ITHE and GMR. In **Fig. 4a**, this parameter is used to study the Pt-thickness dependence while keeping the h-$LuFeO_3$ thickness fixed at 15 nm. ITHE decreases monotonically with increasing Pt thickness, consistent with the interfacial origin: only a few nm of Pt near the interface acquires the topological spin structure from the adjacent h-$LuFeO_3$. In contrast, GMR exhibits a peak value near 5.5 nm. This indicates that, besides the interfacial layer that acquires topological spin structure, an extended layer becomes superparamagnetic, as illustrated in **Fig. 3b**, consistent with the slightly larger cluster size inferred from GMR (52 ± 11 $\mu_B$) compared with ITHE (34 ± 12 $\mu_B$). At very small Pt thickness, the reduced cluster size suppresses GMR, while at large thickness the contribution from the non-superparamagnetic part of Pt dilutes the signal.

Directly probing the magnetic order of h-$LuFeO_3$ has long been challenging in ultrathin films because of its weak magnetization[27]. The MPE-driven ITHE and GMR provides a sensitive alternative probe. In particular, the temperature at which ITHE and GMR vanish, denoted as $T_0$, reflect the onset of long-range magnetic order in h-$LuFeO_3$. The weak dependence of $T_0$ on Pt thickness (**Fig. S9**) confirms that these effects indeed originate from the magnetic ordering of the h-$LuFeO_3$. **Fig. 4b** shows $T_0$ extracted from the temperature dependence of GMR and ITHE in Pt/h-$LuFeO_3$ of different h-$LuFeO_3$ thickness while keeping the Pt thickness fixed at 3.5 nm (see also **Fig. S10**). For thick h-$LuFeO_3$ films, $T_0$ slightly exceeds the bulk $T_N$ (≈ 130 K) of h-$LuFeO_3$[32], possibly due to two-dimensional short-range correlations that persist above the three-dimensional ordering transition[39]. As the h-$LuFeO_3$ thickness decreases, $T_0$ is progressively suppressed, consistent with the finite-size scaling[40–42]. This thickness dependence of the ordering temperature, revealed here down to the ultrathin limit (1.5 unit cells), has not been accessible using conventional magnetometry because of the extremely small magnetization of h-$LuFeO_3$.

## Conclusion

We have established the interfacial topological Hall effect as a sensitive probe of spin topology in insulating magnets, even in the ultrathin limit. Unlike conventional THE, which typically appears as peak-and-dip features near coercivity, the robust noncoplanar spin structure of h-$LuFeO_3$ generates a Hall signal across a broad field range. These findings demonstrate that spin transport couples more strongly to spin topology rather than magnetization and highlight engineered metal/insulator interfaces as promising platforms for exploiting topological spin textures in future spintronic devices.

## Methods

### Film growth and device fabrication

Pt(111) / h-$LuFeO_3$(001), Pt(111) / h-$YbFeO_3$(001), and Pt(111) / h-$LuMnO_3$(001) heterostructures were epitaxially grown on YSZ (111) using pulsed laser deposition (PLD), with a yttrium aluminum garnet (YAG) laser (266-nm wavelength, 70-mJ pulse energy over a spot of ≈2 mm diameter, 3-Hz repetition rate).[31] Thin films of h-$LuFeO_3$, h-$YbFeO_3$, and h-$LuMnO_3$ about 15 nm thick were deposited in 20 mTorr $O_2$ with 650 °C substrate temperature and cooled down in the same pressure to room temperature. After the film-substrate reached room temperature, the chamber was evacuated to $10^{-7}$ Torr vacuum, and the Pt layer of various thickness was deposited *in situ* using the PLD to avoid Pt oxidation and interfacial contamination. The Pt layer was patterned into a Hall bar by photolithography and ion milling.

### Structural characterization

High-resolution x-ray diffraction (XRD) and x-ray reflectivity of epitaxial Pt films were measured using a Rigaku SmartLab Diffractometer with Cu Kα radiation (wavelength 1.54 Å). Crystallinity of oxide and Pt films are studied using XRD. The thickness of the Pt layer is measured using XRR.

### Magneto transport measurements

Magneto-transport measurements were performed in a Quantum Design physical property measurement system equipped with a 9 Tesla longitudinal magnet and horizontal rotator options. Longitudinal ($\rho_{xx}$) and transverse ($\rho_{xy}$) resistivity were measured using the Hall bar configuration [30,31] with the magnetic field applied along different directions at various temperatures using an AC Transport option with 100 μA excitation current at 17 Hz. Field dependence of $\rho_{xx}$ and $\rho_{xy}$ was symmetrized and antisymmetrized[30,31], respectively, to minimize the effects from imperfect sample geometry.

### X-ray absorption and x-ray magnetic circular dichroism

X-ray absorption (XAS) and x-ray magnetic circular dichroism (XMCD) measurements were performed at Beamline 4-ID-D of the Advanced Photon Source at Argonne National Laboratory. The Pt L2 XAS and XMCD data were taken in fluorescence mode using an energy dispersive detector. Samples were measured with magnetic fields up to 600 Oe applied perpendicular to the film plane at temperatures 15 and 280 K and with x-ray of different polarizations. XMCD is calculated using the contrast of XAS measured in the same x-ray polarization but opposite magnetic fields. The Pt L2 XMCD data were taken at 15 K, after field cooling from 280 K in +/-600 Oe applied field along the film normal direction. The incident x-ray direction was a few degrees off from the film normal, probing the net Pt moment along the film normal direction. XMCD is calculated as the contrast in XAS between the two opposite circular polarizations. In order to remove systematic artifacts, the difference between the two field directions is replotted as XMCD.

## Data availability

Source data are provided with this paper. All other data that support the findings of this study are available within the article and Supplementary Information.

## Acknowledgements

This work was primarily supported by the NSF/EPSCoR RII Track-1: Emergent Quantum Materials and Technologies (EQUATE) Award OIA-2044049. (J.L., B.G., Y.Y., X.X.). The work at Fudan University was supported by the National Key Research Program of China 2022YFA1403300 (H. L., W.W., J. S.). Work at Bryn Mawr College (X.W., X.M.C.) is supported by the National Science Foundation (#1708790, #2242796 and #2427091). The transport measurement at Georgia Tech (T.H.Z. and Z.G.J.) was supported by the Department of Energy under grant no. DE-FG02-07ER46451. The work performed at the Advanced Photon Source was supported by the U.S. Department of Energy, Office of Science, and Office of Basic Energy Sciences under Contract No. DE-AC02-06CH11357. The research was performed in part at the Nebraska Nanoscale Facility, National Nanotechnology Coordinated Infrastructure and the Nebraska Center for Materials and Nanoscience, which are supported by the NSF under grant no. ECCS- 2025298, as well as the Nebraska Research Initiative through the Nebraska Center for Materials and Nanoscience and the Nanoengineering Research Core Facility at the University of Nebraska–Lincoln.

## Author contributions

The thin film structures fabrication and structural characterization were carried out by J.L. and H.L. with assistance from W.W., Y.Y., S. J., and X.X. Transport measurements were carried out by A.H.C, A.M, T.Z, and H.L., with assistance from J.L., D.S., W.W., Z.J., and X.X. X-ray absorption spectroscopy was studied by X.W. X.C, and Y.C. The study was conceived by D.S., W.W., and X.X. J.L. and X.X. co-wrote the manuscript. All authors discussed the results and commented on the manuscript.

## Competing interests

The authors declare no competing interests.

## References

1. Xiao, D., Chang, M. C. & Niu, Q. Berry phase effects on electronic properties. *Reviews of Modern Physics* **82**, 1959–2007 (2010).

2. Everschor-Sitte, K. & Sitte, M. Real-space Berry phases: Skyrmion soccer (invited). *Journal of Applied Physics* **115**, 172602 (2014).

3. Bruno, P., Dugaev, V. K. & Taillefumier, M. Topological Hall effect and Berry phase in magnetic nanostructures. *Physical Review Letters* **93**, 096806 (2003).

4. Liang, X. *et al.* Influence of Interface Structure on Magnetic Proximity Effect in Pt/$Y_3Fe_5O_{12}$ Heterostructures. *ACS Appl. Mater. Interfaces* **8**, 8175–8183 (2016).

5. Amamou, W. *et al.* Magnetic proximity effect in Pt/CoFe2O4 bilayers. *Phys. Rev. Materials* **2**, 011401(R) (2018).

6. Hauser, J. J. Magnetic Proximity Effect. *Phys. Rev.* **187**, 580–583 (1969).

7. Bhattacharyya, S. *et al.* Recent Progress in Proximity Coupling of Magnetism to Topological Insulators. *Advanced Materials* **33**, 2007795 (2021).

8. Huang, S. X. & Chien, C. L. Extended Skyrmion Phase in Epitaxial FeGe(111) Thin Films. *Physical Review Letters* **108**, 267201 (2012).

9. Neubauer, A. *et al.* Topological Hall Effect in the A Phase of MnSi. *Physical Review Letters* **102**, 186602 (2009).

10. Kanazawa, N. *et al.* Large Topological Hall Effect in a Short-Period Helimagnet MnGe. *Physical Review Letters* **106**, 156603 (2011).

11. Takagi, H. *et al.* Spontaneous topological Hall effect induced by non-coplanar antiferromagnetic order in intercalated van der Waals materials. *Nat. Phys.* **19**, 961–968 (2023).

12. Song, A. *et al.* Large Anomalous Hall Effect in a Noncoplanar Magnetic Heterostructure. *Adv Funct Materials* **35**, 2422040 (2025).

13. Bauer, J. J. *et al.* Magnetic proximity effect in magnetic-insulator/heavy-metal heterostructures across the compensation temperature. *Phys. Rev. B* **104**, 094403 (2021).

14. Geprägs, S. *et al.* Static magnetic proximity effects and spin Hall magnetoresistance in Pt/Y3Fe5O12 and inverted Y3Fe5O12/Pt bilayers. *Phys. Rev. B* **102**, 2469 (2020).

15. Geprägs, S. *et al.* Investigation of induced Pt magnetic polarization in Pt/Y3Fe5O12 bilayers. *Applied Physics Letters* **101**, 262407 (2012).

16. Lu, Y. M. *et al.* Pt Magnetic Polarization on Y3Fe5O12 and Magnetotransport Characteristics. *Physical Review Letters* **110**, 147207 (2013).

17. Jungfleisch, M. B., Lauer, V., Neb, R., Chumak, A. V. & Hillebrands, B. Improvement of the yttrium iron garnet/platinum interface for spin pumping-based applications. *Applied Physics Letters* **103**, 022411 (2013).

18. Huang, S. Y. *et al.* Transport Magnetic Proximity Effects in Platinum. *Physical Review Letters* **109**, 107204 (2012).

19. Shiomi, Y. *et al.* Interface-dependent magnetotransport properties for thin Pt films on ferrimagnetic Y3Fe5O12. *Applied Physics Letters* **104**, 242406 (2014).

20. Meyer, S. *et al.* Anomalous Hall effect in YIG|Pt bilayers. *Applied Physics Letters* **106**, 132402 (2015).

21. Shimizu, S. *et al.* Electrically Tunable Anomalous Hall Effect in Pt Thin Films. *Phys. Rev. Lett.* **111**, 216803 (2013).

22. Miao, B. F., Huang, S. Y., Qu, D. & Chien, C. L. Physical Origins of the New Magnetoresistance in Pt/YIG. *Physical Review Letters* **112**, 236601 (2014).

23. Li, X., Yun, Y. & Xu, X. Recent progress on multiferroic hexagonal rare-earth ferrites (h-RFeO3, R = Y, Dy-Lu). *Journal of Physics D: Applied Physics* **58**, 073003 (2025).
24. Xu, X. & Wang, W. Multiferroic hexagonal ferrites (h-RFeO3,R = Y, Dy-Lu): a brief experimental review. *Modern Physics Letters B* **28**, 1430008 (2014).
25. Das, H., Wysocki, A. L., Geng, Y., Wu, W. & Fennie, C. J. Bulk magnetoelectricity in the hexagonal manganites and ferrites. *Nature Communications* **5**, 2998 (2014).
26. Wang, W. *et al.* Room-Temperature Multiferroic Hexagonal LuFeO_{3} Films. *Physical Review Letters* **110**, 237601 (2013).
27. Moyer, J. A. *et al.* Intrinsic magnetic properties of hexagonal LuFeO3 and the effects of nonstoichiometry. *APL Materials* **2**, 012106 (2014).
28. Kimbell, G., Kim, C., Wu, W., Cuoco, M. & Robinson, J. W. A. Challenges in identifying chiral spin textures via the topological Hall effect. *Commun Mater* **3**, 19 (2022).
29. Solovyev, I. V., Valentyuk, M. V. & Mazurenko, V. V. Magnetic structure of hexagonal YMnO3and LuMnO3from a microscopic point of view. *Phys. Rev. B* **86**, 054407 (2012).
30. Li, J., Comstock, A. H., Sun, D. & Xu, X. Comprehensive demonstration of spin Hall Hanle effects in epitaxial Pt thin films. *Physical Review B* **106**, 184420 (2022).
31. Li, J. *et al.* Giant interfacial spin Hall angle from Rashba-Edelstein effect revealed by the spin Hall Hanle process. *Physical Review B* **108**, L241403 (2023).
32. Sinha, K. *et al.* Tuning the N\'eel Temperature of Hexagonal Ferrites by Structural Distortion. *Physical Review Letters* **121**, 237203 (2018).
33. Hansen, P., Röschmann, P. & Tolksdorf, W. Saturation magnetization of gallium-substituted yttrium iron garnet. *Journal of Applied Physics* **45**, 2728–2732 (1974).

34. Kim, K. *et al.* Large anomalous Hall current induced by topological nodal lines in a ferromagnetic van der Waals semimetal. *Nature Mater* **17**, 794–799 (2018).

35. Cao, S. *et al.* Electronic structure and direct observation of ferrimagnetism in multiferroic hexagonal $YbFeO_3$. *Physical Review B* **95**, 224428 (2017).

36. Jeong, Y. K. *et al.* Structurally Tailored Hexagonal Ferroelectricity and Multiferroism in Epitaxial YbFeO3 Thin-Film Heterostructures. *Journal of the American Chemical Society* **134**, 1450–1453 (2012).

37. Zhang, T. Z. *et al.* Spin Hall magnetoresistance in antiferromagnetic $\alpha$ -Fe2O3/Pt bilayers: Modulation from interface magnetic state. *Applied Physics Letters* **121**, 262404 (2022).

38. Baibich, M. N. *et al.* Giant magnetoresistance of (001)Fe/(001)Cr magnetic superlattices. *Physical Review Letters* **61**, 2472–2475 (1988).

39. Lonkai, Th. *et al.* Magnetic two-dimensional short-range order in hexagonal manganites. *Journal of Applied Physics* **93**, 8191–8193 (2003).

40. Zhang, R. & Willis, R. F. Thickness-dependent curie temperatures of ultrathin magnetic films: Effect of the range of spin-spin interactions. *Physical Review Letters* **86**, 2665–2668 (2001).

41. Fisher, M. E. & Barber, M. N. Scaling Theory for Finite-Size Effects in the Critical Region. *Physical Review Letters* **28**, 1516–1519 (1972).

42. Vaz, C. A. F., Bland, J. A. C. & Lauhoff, G. Magnetism in ultrathin film structures. *Reports on Progress in Physics* **71**, 56501 (2008).

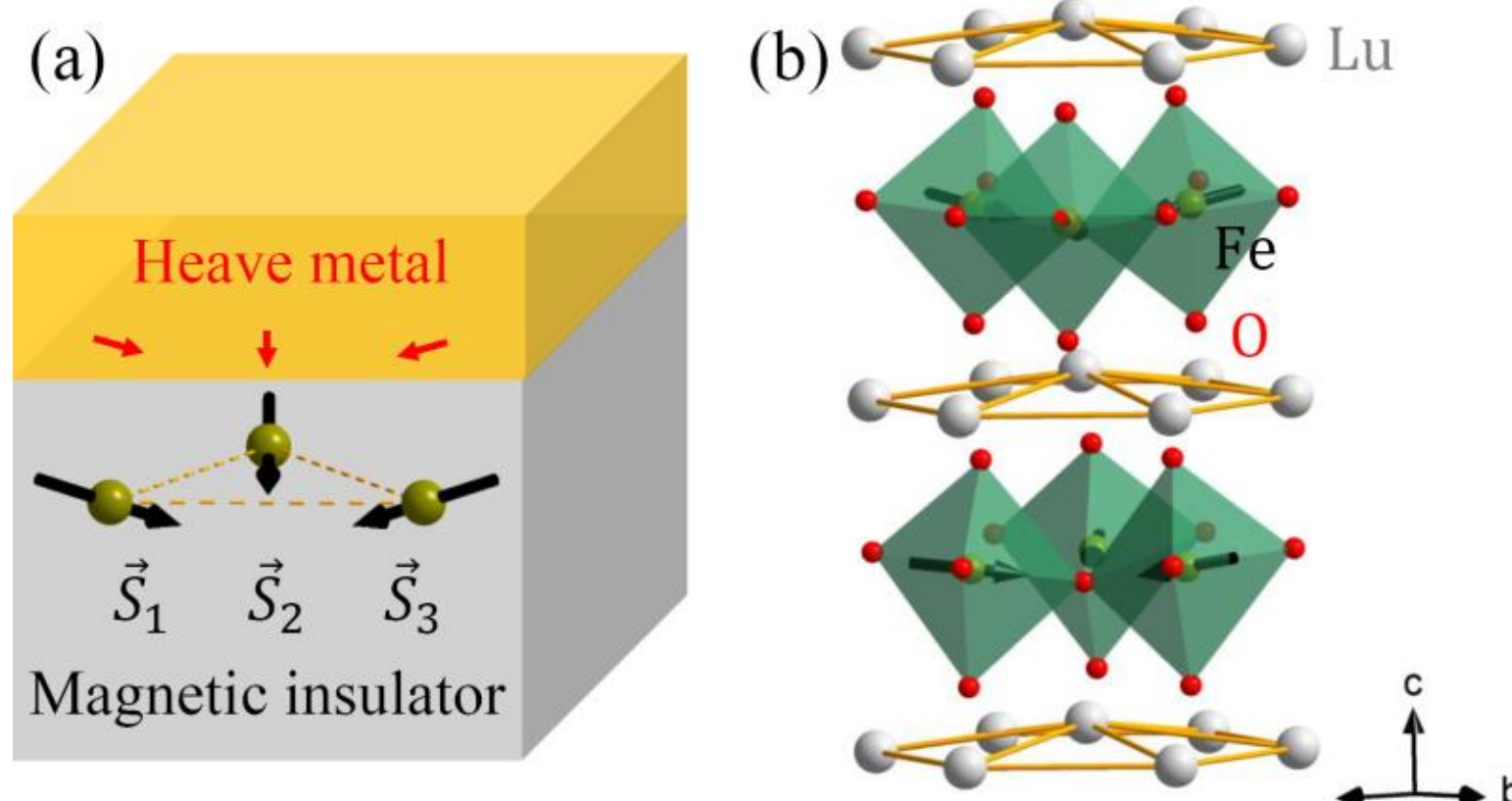

**Figure 1. Induced topological spin structure in heavy metal via magnetic proximity effect (MPE)** (a) Schematic of a bilayer system in which three spins (**S**$_1$, **S**$_2$, **S**$_3$) in the magnetic insulator form a non-coplanar trimer of non-trivial topology. This spin topology is transferred to the adjacent heavy metal via the MPE. (b) Crystal structure of hexagonal $LuFeO_3$ (h-$LuFeO_3$), where Fe spins realize the non-coplanar trimer arrangement in (a).

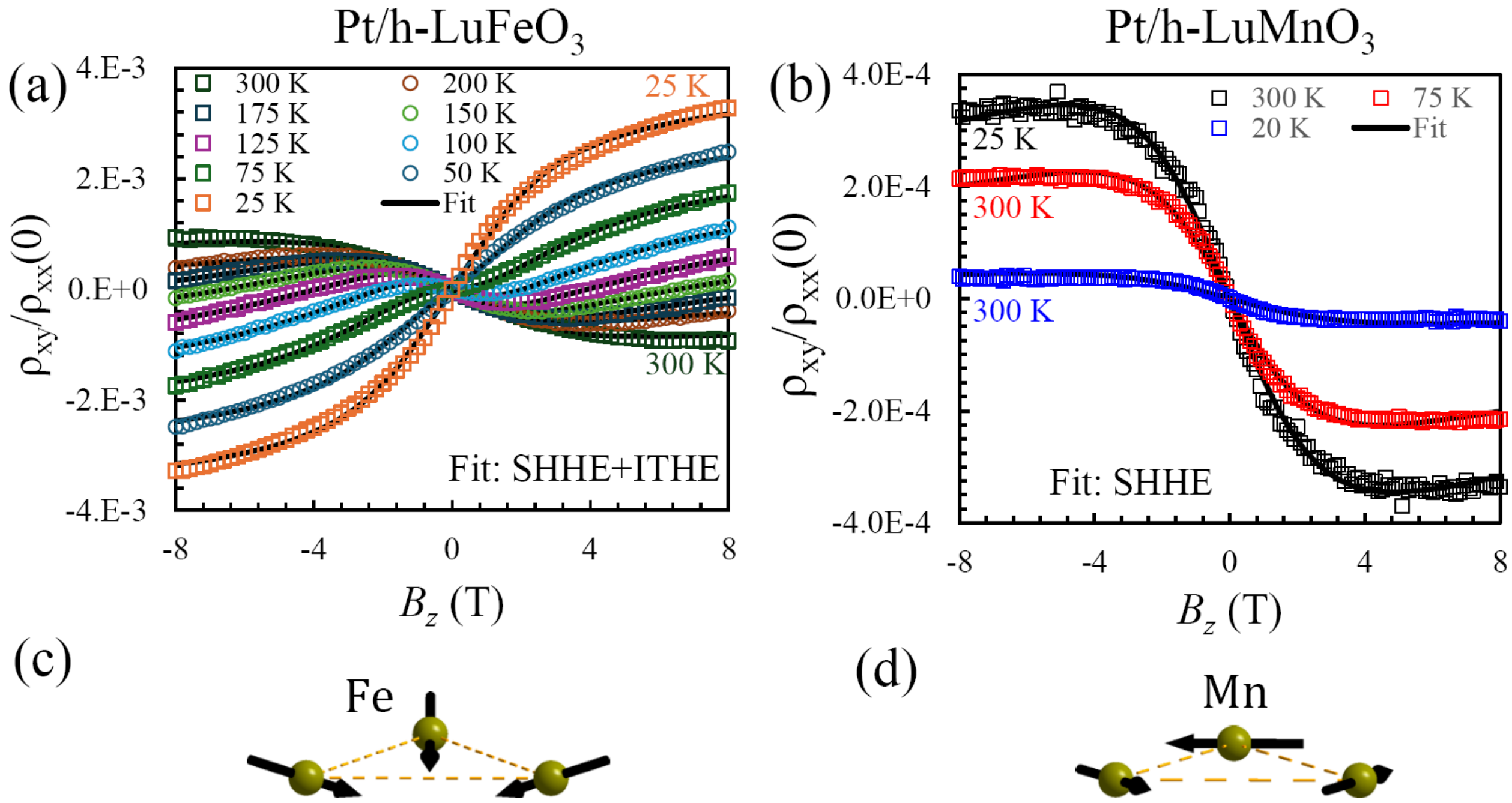

**Figure 2. Interfacial topological Hall effect (ITHE) in Pt/h-$LuFeO_3$.** (a) Hall resistivity normalized by longitudinal resistivity, $\rho_{xy}(B_z)/\rho_{xx}(0)$, for a Pt (3.1 nm)/ h-$LuFeO_3$ (15 nm) bilayer. The signal is dominated by a negative component (SHHE) at room temperature but switches to a positive component (ITHE) at low temperature when the magnetic order develops below $T_N \approx 130$ K. (b) In contrast, $\rho_{xy}(B_z)/\rho_{xx}(0)$ in a Pt (3.5 nm)/ $LuMnO_3$ (35 nm) bilayer remains negative (SHHE) across all temperatures. (c-d) The essential difference is in spin topology: Fe spin trimers in h-$LuFeO_3$ form a non-coplanar configuration with non-trivial topology, generating an emergent magnetic field responsible for ITHE. By contrast, Mn spin trimers in h-$LuMnO_3$ form a coplanar structure with trivial topology, producing no emergent field and hence no ITHE.

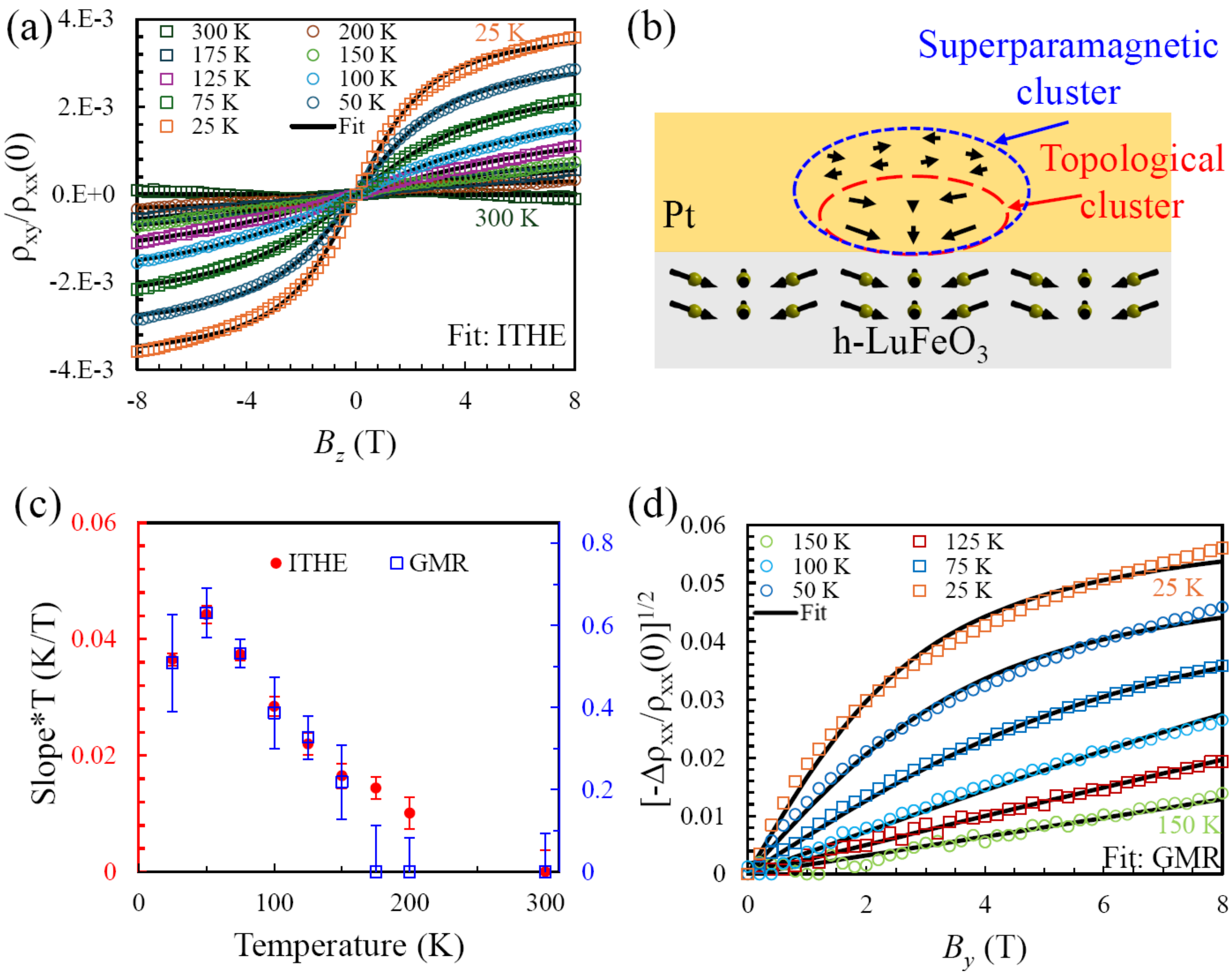

**Figure 3**. **Interfacial topological Hall effect (ITHE) and giant magnetoresistance (GMR) in the Pt (3.1 nm)/h-LuFeO$_3$ (15 nm) sample.** (a) Extracted ITHE contribution $\rho_{xy,ITHE} = \rho_{xy}-\rho_{xy,SHHE}$ normalized by $\rho_{xx}(0)$ together with fits based on a superparamagnetic nanocluster model. (b) Schematic illustration of Pt nanoclusters acquiring non-trivial topology via magnetic proximity to h-LuFeO$_3$. The interfacial Pt layer inherits the spin chirality from Fe trimers. (c) Temperature dependence of the order parameter, defined as the low-field slope in (a) or (d) multiplied by temperature (see text). (d) Square root of the negative magnetoresistance $[-\Delta\rho_{xx}(B_y)/\rho_{xx}(0)]^{1/2}$ showing similar field and temperature dependence as ITHE in (a), with fits based on GMR from superparamagnetic nanoclusters.

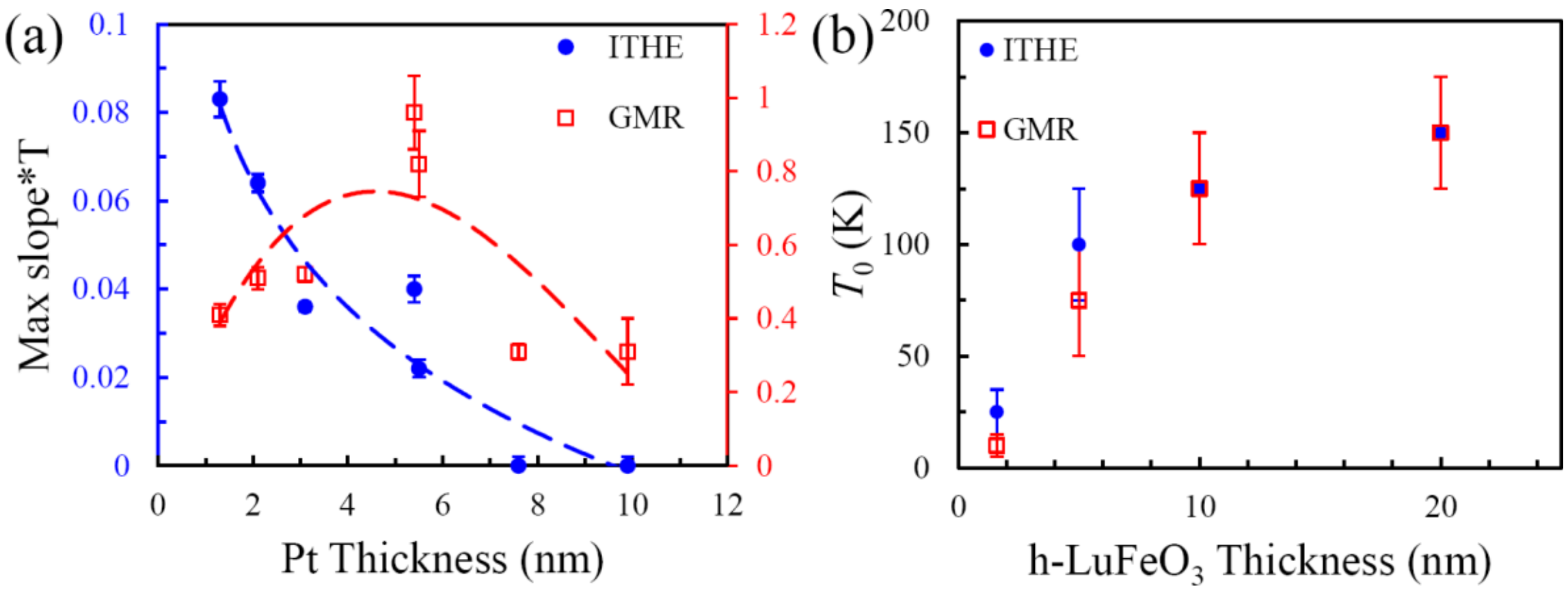

**Figure 4. Thickness dependence of ITHE and GMR.** (a) Pt-thickness dependence of the magnetic order parameter extracted from ITHE and GMR. Dashed lines are guide to the eye. (b) h-$LuFeO_3$-thickness dependence of $T_0$, defined as the temperature at which ITHE and GMR vanish.